\documentclass[prl,
preprintnumbers,amsmath,amssymb]{revtex4}
\usepackage{graphicx}

\DeclareGraphicsExtensions{.pdf,*.jpg}

 \def\be{\begin{equation}}
 \def\ee{\end{equation}}

 \def\A{\mathcal{A}}

 \def\2{\frac{1}{2}}
 \def\4{\frac{1}{4}}

\catcode`\@=11
\def\@citex[#1]#2{%
\if@filesw \immediate \write \@auxout {\string \citation {#2}}\fi
\@tempcntb\m@ne \let\@h@ld\relax \def\@citea{}%
\@cite{%
  \@for \@citeb:=#2\do {%
    \@ifundefined {b@\@citeb}%
      {\@h@ld\@citea\@tempcntb\m@ne{\bf ?}%
      \@warning {Citation `\@citeb ' on page \thepage \space
undefined}}%
      {\@tempcnta\@tempcntb \advance\@tempcnta\@ne%
      \@tempcntb\number\csname b@\@citeb \endcsname \relax%
      \ifnum\@tempcnta=\@tempcntb %Number follows previous--hold on to
it
        \ifx\@h@ld\relax%
          \edef \@h@ld{\@citea\csname b@\@citeb\endcsname}%
        \else%
          \edef\@h@ld{\ifmmode{-}\else--\fi\csname
b@\@citeb\endcsname}%
        \fi%
      \else%   %  non-successor--dump what's held and do this one
        \@h@ld\@citea\csname b@\@citeb \endcsname%
        \let\@h@ld\relax%
      \fi}%
    \def\@citea{,\penalty\@highpenalty\,}%
  }\@h@ld
}{#1}}

\def\@citeb#1#2{{[#1]\if@tempswa , #2\fi}}
\def\@citeu#1#2{{$^{#1}$\if@tempswa , #2\fi }}
\def\@citep#1#2{{#1\if@tempswa , #2\fi}}

\begin{document}
\preprint{CERN-TH-2019-177}

\title{Corrections to de Sitter entropy through holography}

\author{N. Tetradis}
\email{ntetrad@phys.uoa.gr}
\affiliation{%
Department of Physics,
University of Athens,
University Campus,
Zographou 157 84, Greece
\\ and 
\\
Theoretical Physics Department, 
CERN, CH-1211 
Geneva 23, Switzerland}
\date{\today}%

\begin{abstract}
The holographic entanglement entropy is computed for an entangling
surface that coincides with the horizon of a boundary de Sitter  
metric. This is achieved through an appropriate slicing of anti-de Sitter 
space and the implementation of a UV cutoff.
The entropy is equal to the Wald
entropy for an effective action that includes the higher-curvature 
terms associated with the conformal anomaly. 
The UV cutoff can be expressed in terms of the effective Planck mass 
and the number of degrees of freedom of the dual theory.
The entanglement entropy takes the expected form of the de Sitter entropy,
including logarithmic corrections.

\end{abstract}

%\pacs{}
% PACS, the Physics and Astronomy Classification Scheme.
\maketitle

%\section{Introduction}\label{intro}

The relation between 
entanglement and gravitational entropy in spaces that contain 
horizons can shed light to the nature of the latter.
The divergent part of the entanglement entropy scales with the area 
of the entangling surface $\A$. This feature 
hints at a connection with the gravitational entropy when this 
surface is identified with a horizon \cite{sorkin}.
In the context of the
AdS/CFT correspondence \cite{adscft}, the conjecture of \cite{ryu,review} 
states that the entanglement entropy is proportional to the
area of a minimal surface 
$\gamma_A$ extending from $\A$ into the bulk. 
We apply here this conjecture to anti-de Sitter (AdS) space  
with a boundary de Sitter (dS) metric  \cite{giatet}.

We consider the 
slicing of AdS space that results in a metric 
with a static dS boundary:
\begin{equation}
ds^2_{d+2}
= \frac{R^2}{z^2} \left[ dz^2 +\left(1-\frac{1}{4}H^2 z^2 \right)^2 \left(
- (1-H^2\rho^2) dt^2 
+  \frac{d\rho^2}{1-H^2\rho^2}+\rho^2 \, d\Omega^2_{d-1} \right) \right],
\label{dS} \end{equation}
where $0\leq \rho \leq 1/H$ covers a static patch for $d>1$. There are two 
such patches in the global geometry, 
with $\rho=0$ corresponding to the ``North" and ``South poles".
For $d=1$, $\rho$ can be negative and each static
patch is covered by $-1/H \leq \rho \leq 1/H$. 
All the coordinates
in the above expressions, as well as $H$, are taken to be dimensionless,
with $R$ the only dimensionful parameter. The dimensionality of the various
quantities can be reinstated by multiplication with the appropriate powers of $R$. 
In particular, the
physical Hubble scale is $H/R$.
Eq. (\ref{dS}) is a particular example of a  
Fefferman-Graham parametrization of AdS space \cite{fg}.

\begin{figure}[t]
\centering
\includegraphics[width=0.5\textwidth]{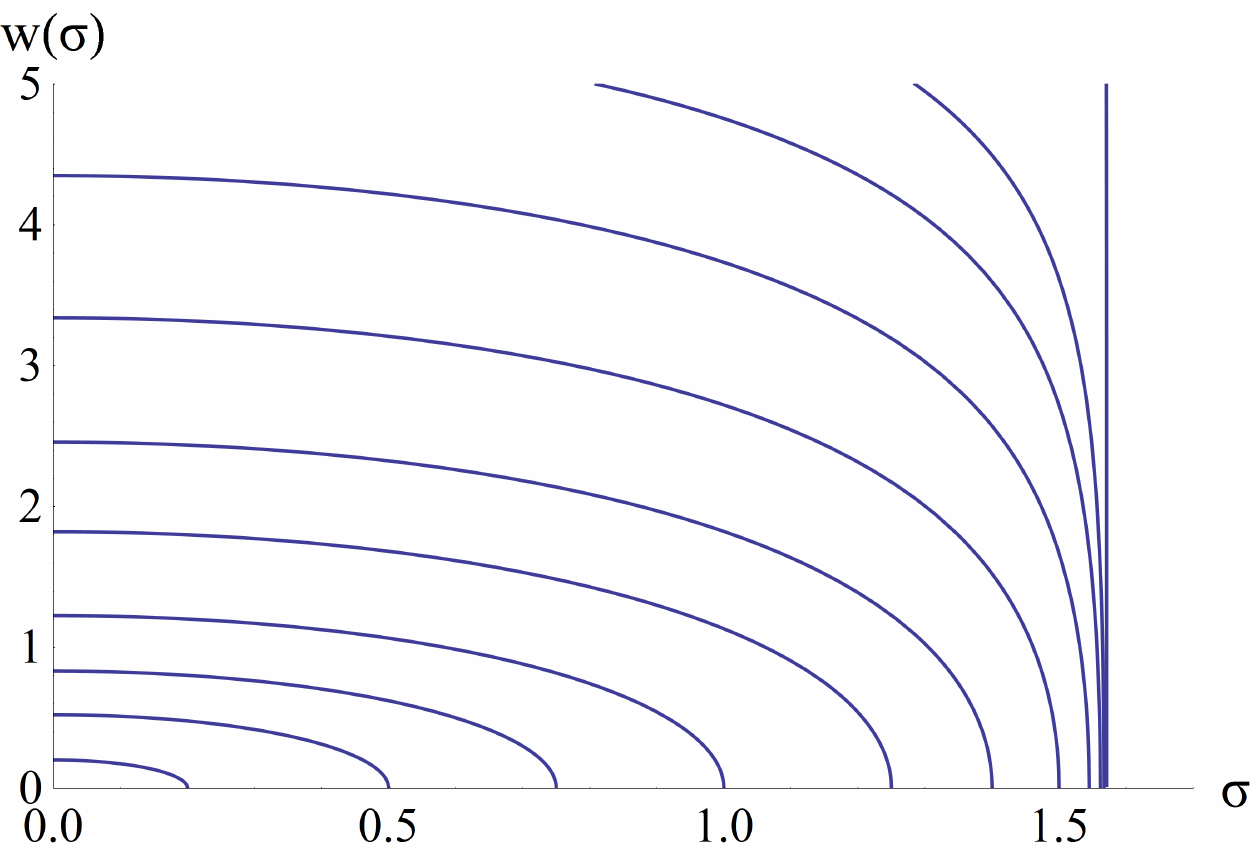}
\caption{Minimal surfaces for a de Sitter boundary.}
\label{solution}
\end{figure}

The minimal surface $\gamma_A$ in the bulk can be determined through the minimization of the area
\be
{\rm Area}(\gamma_A)=R^d S^{d-1}\int d\rho\, \rho^{d-1}
\frac{\left(1-\frac{1}{4}H^2z^2\right)^{d-1}}{z^d}
\sqrt{\frac{\left(1-\frac{1}{4}H^2z^2\right)^{2}}{1-H^2\rho^2}
+\left(\frac{dz(\rho)}{d\rho}\right)^2  },
\label{areads} \ee
with $S^{d-1}$ the volume of the ($d-1$)-dimensional unit sphere.
Through the definitions $\sigma=\sin^{-1}(H\rho)$, $w=2\tanh^{-1}(Hz/2)$, the 
above expression becomes
\be
{\rm Area}(\gamma_A)=R^d S^{d-1}\int 
d\sigma \frac{\sin^{d-1}(\sigma)}{\sinh^{d}(w)}\sqrt{1+\left(\frac{dw(\sigma)}{d\sigma}\right)^2}.
\label{areads2} \ee
Minimization of the area results in the differential equation
\be
\tan(\sigma)\tanh(w)\,w''+(d-1)\tanh(w)\left(\left(w'\right)^3+w'\right)
+d\tan(\sigma)\left(\left(w'\right)^2+1 \right)=0,
\label{diffds} \ee
whose solution is
\be
w(\sigma)=\cosh^{-1} \left( 
 \frac{\cos(\sigma)}{\cos(\sigma_0)}\right).
\label{solws} \ee 
%\be
%w(\sigma)=2 \,{\rm arctanh} \left( 
%\sqrt{\frac{2+\cos(2\sigma)-4\cos(\sigma)\cos(\sigma_0)+\cos(2\sigma_0)}{\cos(2\sigma)-\cos(2\sigma_0)} } \right).
%\label{solws} \ee
%Its inverse is 
%\be
%\sigma(w)=\arccos \left( \cos(\sigma_0) \cosh(w) \right).
%\label{inverse} \ee
 For $\sigma_0\to 0$ the solution becomes
$w(\sigma)=\sqrt{\sigma_0^2-\sigma^2}$, reproducing the known expression for 
$H=0$ \cite{ryu,review}. For $\sigma_0\to \pi/2$ 
the solution approaches the boundary at the location of the 
horizon  with $dw/d\sigma \to -\infty$. We also have 
$w(0)\to \infty$ in this limit.
In fig. \ref{solution} we depict the solution for 
increasing values of $\sigma_0$.
The solution for a minimal surface in AdS space with dS slicing has also been 
derived in \cite{grieninger} in the context of a computation of
entanglement entropy on finite-cutoff surfaces. 

The total area of the minimal surface is dominated by the region near the boundary. 
Cutting off the range of $z$ at $z=\epsilon$, gives for the 
divergent part
\be
{\rm Area}(\gamma_A)=
R^d S^{d-1}I(\epsilon) =
R^d S^{d-1}\int_{H\epsilon} 
\frac{dw}{\sinh^{d}(w)}.
\label{areaeps} \ee 
The leading contribution to $I(\epsilon)$ is 
$I(\epsilon)=1/((d-1)H^{d-1}\epsilon^{d-1})$ for $d\not= 1$, and
 $\log(1/ (H \epsilon$)) for $d=1$.
In \cite{giatet} it was argued that a connection with the 
entropy associated with a gravitational background can be established 
if we define the effective
Newton's constant for the boundary theory following \cite{hawking}:
\be
G_{d+1}=(d-1) \epsilon^{d-1}\frac{G_{d+2}  }{R},
\label{Geff} \ee
with $(d-1) \epsilon^{d-1}$ replaced by $1/\log(1/\epsilon)$ for $d=1$.
This definition is natural
within an effective theory that implements consistently a cutoff procedure
by eliminating the part of AdS space corresponding to $z<\epsilon$. Such
a framework is provided, for example, by the Randall-Sundrum (RS) model \cite{rs}.
 As pointed out in \cite{hawking}, 
the cutoff dependence of the four-dimensional Newton's constant is not apparent 
in \cite{rs} because the metric is rescaled by $\epsilon^2$.

In order to see how eq. (\ref{Geff}) may arise, we consider the 
AdS parametrization used in \cite{rs}:
\be
ds^2_{d+2}
= e^{-2k r_c \phi} \eta_{\mu\nu}dx^\mu dx^\nu+r_c^2 d\phi^2.
\label{dsmetric} \ee
It corresponds to a flat slicing of AdS with a Fefferman-Graham 
coordinate $z=\exp(kr_c \phi)$ and AdS radius $R=1/k$. The parameter
$r_c$ determines the distance between the positive-tension brane 
located at $\phi=0$ and the negative-tension brane at $\phi=\pi$.
The orbifold construction assumes a reflection symmetry around $\phi=0$,
or $z=1$,
with an identification of the points $\phi=\pm \pi$.
Taking the limit $r_c\to\infty$ removes the negative-tension brane.
Gravitational excitations are considered by taking 
$\eta_{\mu\nu}\to \eta_{\mu\nu}+h_{\mu\nu}$ and treating $h_{\mu\nu}$
as a perturbation. 
The curvature scalar of the background 
has a $\delta$-function singularity at $\phi=0$,
which leads to the localization of a massless excitation mode 
around this point. The effective Newton's constant for the
low-energy theory is 
\be
G_{d+1}^{-1} =2 G_{d+2}^{-1} \int_0^{r_c\pi} dy\, e^{-(d-1) k y}
=\frac{2}{(d-1)k}G_{d+2}^{-1},
\label{Geffrs} \ee
for $r_c\to \infty$,
with the factor of 2 coming from the two copies of AdS in the construction.
In order to reproduce eq. (\ref{Geff}) we need to place the positive
tension brane at $\phi=-\pi$, the negative tension brane at $\phi=0$, and
identify the points $\phi=-2\pi$ and 0. The limit $r_c\to\infty$ now takes
the positive tension brane arbitrarily close to the AdS boundary and
results in an effective Newton's constant given by twice the value 
of eq. (\ref{Geff}) with
$\epsilon=\exp(-k r_c\pi)$. The factor of 2 in eq. (\ref{Geffrs}) does not
appear in (\ref{Geff}) because the latter accounts for only one copy
of AdS space, as opposed to two in the RS scenario. A similar factor of 2
would appear in the total entropy, with the two factors cancelling out
in the final expression. We shall carry out our analysis based on 
eq. (\ref{Geff}) in order to be consistent with holographic renormalization.

%This definition can also be viewed
%as resulting from the regulated action in the 
%context of holographic renormalization \cite{skenderis,papadimitriou}.

The use of eq. (\ref{Geff}) gives a 
leading contribution to the entropy 
\be
S_{\rm dS}=\frac{{\rm Area}(\gamma_A)}{4G_{d+2}}=
\frac{R^d S^{d-1}}{4G_{d+2}(d-1)H^{d-1}\epsilon^{d-1}}=
\frac{S^{d-1}}{4G_{d+1}}\left(\frac{R}{H}\right)^{d-1}
=\frac{A_H}{4G_{d+1}},
\label{entropydss} \ee
with $A_H$ the area of the horizon. This expression
reproduces the gravitational entropy of \cite{gibbons} 
and agrees with the findings of \cite{nguyen}.
Notice that, according to our conventions, the physical
Hubble scale is $H/R$.
The result is valid for $d=1$ as well, 
with $1/((d-1)\epsilon^{d-1})$ replaced by $\log(1/ \epsilon$)
and $S^0=2$, because the horizons of the global dS$_2$ geometry consist of 2 points
\cite{hawking}.

We turn next to the 
subleading divergences that may appear in the integral $I(\epsilon)$ 
for $\epsilon\to 0$. 
Even though we focus here on the dominant terms, 
it must be pointed out that the solution (\ref{solws}) is
exact, so that, in principle, the entropy can be computed exactly through
eq. (\ref{areads2}), including the finite terms. 
For $d=3$ we have 
\be
I(\epsilon)=\frac{1}{2H^2\epsilon^2}+\frac{1}{2}\log(H \epsilon)+{\cal O}(\epsilon^0),
\label{expandI} \ee
for $d=2$ there are no singular subleading terms, while for 
$d=1$ the logarithmic term, already accounted for in eq. (\ref{entropydss}), is the leading divergence with no subleading ones.
For $d>3$ we obtain subleading power-law divergences for odd $d+1$, augmented
by a logarithmic one for even $d+1$.
The implication of eq. (\ref{expandI})
is that the dS entropy in four dimensions is proportional to the area of the horizon,
with a coefficient that receives a logarithmic correction:
\be
S_{\rm dS}=\frac{A_H}{4G_4}\left(
1+H^2 \epsilon^2 \log H \epsilon \right),
\label{holentropydS} \ee
where we have made use of eq. (\ref{Geff}).

%It is an interesting 
%problem to identify the origin of the correction. It 

The correction must be attributed to 
higher-curvature terms in the effective gravitational action.
For a four-dimensional boundary theory ($d=3$), on which we focus, the logarithmic 
dependence on the cutoff implies a connection with the conformal anomaly of
the dual theory. 
The most straightforward way to obtain the effective action for our setup is
through known results in the context of holographic renormalization
\cite{hol1,skenderis,papadimitriou}. The bulk metric of a five-dimensional
asymptotically AdS space is written in a
Fefferman-Graham expansion \cite{fg} as
\begin{eqnarray}
ds^2&=&\frac{R^2}{z^2}\left( dz^2+g_{ij}(x,z)dx^idx^j\right)
\nonumber \\
g(x,z)&=&g_{(0)}+z^2g_{(2)}+z^4g_{(4)}+z^4\log z^2 h_{(4)} 
+{\cal O}(z^5).
\label{fgg} \end{eqnarray}
A solution is then obtained order by
order. The on-shell gravitational action is 
regulated by restricting the bulk integral to the region $z>\epsilon$.
The divergent terms are subtracted through the introduction of 
appropriate counterterms. In this way a renormalized effective action
is obtained, expressed in terms of the induced metric 
$\gamma_{ij}$ on the surface at $z=\epsilon$.  

The holographic entanglement entropy displays
divergences similar to those of the on-shell action 
for $\epsilon\to 0$. In our approach the entropy is not
renormalized. We assume that the cutoff $\epsilon$ is physical and we
incorporate it in the effective couplings.
This amounts to employing the regulated form of the effective action,
without the subtraction of divergences. 
The leading terms, which would diverge for $\epsilon \to 0$, 
can be found in the counterterm action of
holographic renormalization. They are expressed in terms of the induced metric
$\gamma_{ij}$. 
It is apparent from eqs. (\ref{fgg}) that $\gamma_{ij}$ on
a surface at $z=\epsilon$ includes a factor
$\epsilon^{-2}$. In the RS model this factor is displayed explicitly and 
determines
the relative size of the couplings of the effective theory at various locations of
the brane.
Using the results of \cite{skenderis} and extracting the $\epsilon^{-2}$ factor
from $\gamma_{ij}$, we can express the leading terms of
the regulated action as 
\be
S=\frac{R^3}{16\pi G_5}\int d^4x\, \sqrt{-\gamma}
\left[ \frac{6}{\epsilon^4}+\frac{1}{2\epsilon^2}{\cal R}
-\frac{1}{4}\log\epsilon 
\left( {\cal R}_{ij}{\cal R}^{ij}-\frac{1}{3}{\cal R}^2
\right) \right],
\label{effaction} \ee
where we have multiplied with an appropriate power of 
$R$ so that all quantities within the integral are dimensionless, in agreement
with our earlier conventions. We have also 
adapted the results of \cite{skenderis} to a metric of Lorentzian signature.
The first term corresponds to a cosmological constant, which must be
(partially) cancelled by vacuum energy localized on the surface at $z=\epsilon$,
such as the brane tension in the RS model \cite{rs}.
The second term is the standard Einstein term if the effective Newton's
constant $G_4$ is defined according to eq. (\ref{Geff}) with $d=3$. 
The third term is responsible for the holographic conformal anomaly.
The structure of the effective action is the same for the standard RS model
in bulk space with Einstein gravity,
apart from rescalings that set the location of the brane at a 
finite distance equal to $R$
from the boundary \cite{kiritsisRS,myers2}.  

Assuming that the cosmological constant can be adjusted to the
desired value, the action (\ref{effaction}) supports a dS solution.
This can be checked explicitly through the Einstein equations, in
which the contribution from the anomaly term vanishes for a 
dS background. 
%Alternatively one may observe that the metric
%(\ref{dS}) with $d=3$ is in the Fefferman-Graham form 
%(\ref{fgg}) with $h_{(4)}=0$. As $h_{(4)}$ is proportional to the variation
%of the anomaly term with respect to $\gamma_{ij}$  \cite{skenderis}, 
%this variation must vanish for a dS boundary. 
The gravitational entropy must take into account the
presence of the third term in eq. (\ref{effaction}). This can be achieved
by computing the Wald entropy \cite{wald}, which gives the horizon entropy
in theories with higher curvature interactions. 
For the action (\ref{effaction}) the Wald entropy reduces to 
\be
S_{\rm Wald}=\frac{A_H}{4G_4}-\frac{R^3}{32G_5} \log \epsilon
\int_{\cal A} d^2y \sqrt{h}\left( 2 {\cal R}^{ij}\gamma_{ij}^\perp
-\frac{4}{3} {\cal R} \right),
\label{waldentropy} \ee
where the integration is over the horizon, with induced metric $h$, 
and $\gamma^\perp$ denotes the
metric in the transverse space. 
We have made use of eq. (\ref{Geff}) and of the fact that the physical
Hubble scale is $H/R$.
This expression has also been derived in \cite{solodukhin} in the context of
a holographic calculation of the black-hole entropy.
It is in qualitative agreement with \cite{myers2}, even though the logarithmic
term is replaced by a constant there, 
as the RS brane is located at a finite distance from
the boundary. 
For a dS background the term in the parenthesis
in the r.h.s. is constant and the integration reproduces the area of the 
horizon. We obtain
\be
S_{\rm Wald}=\frac{A_H}{4G_4}\left(
1+H^2 \epsilon^2 \log  \epsilon \right).
\label{waldentropydS} \ee
The correction to the dS entropy in eq. (\ref{waldentropydS}) 
is in agreement with the singular part of the
correction provided by the holographic calculation (\ref{holentropydS}).
%The breaking of the conformal invariance by the cutoff $\epsilon$ through
%the anomaly  is reflected in the correction to the dS entropy.

The logarithmic term in eq. (\ref{holentropydS}) can be compared with
the known logarithmic correction to the black-hole entropy \cite{sen}, which
is of the form $S_{\rm bh}=S_{\rm BH}+C \log a,$
where $S_{\rm BH}$ is the Bekenstein-Hawking entropy \cite{bekhawk} and
$a$ the black-hole size parameter, such that the horizon area scales $\sim a^2$
in four dimensions. The constant $C$ is related to the conformal anomaly and
depends on the number of massless 
scalar, Dirac and vectors fields of the theory.
%Its precise value depends on the gravitational background, 
%so that a quantitative comparison with \cite{sen} is not possible.
%On the other hand,   
The field content of the ${\cal N}=4$ supersymmetric $SU(N)$ gauge 
theory in the large-$N$ limit includes $n_S=6N^2$ scalars, $n_F=2N^2$ Dirac fermions and $n_V=N^2$ vectors. 
The Weyl-squared and Euler-density terms in the divergent part of the  
effective action have opposite coefficients and combine into a term \cite{bd}
\be
S=-\frac{\beta}{16\pi^2} \Gamma\left(2-\frac{d+1}{2}\right) 
\int d^4x\, \sqrt{-\gamma}
\left( {\cal R}_{ij}{\cal R}^{ij}-\frac{1}{3}{\cal R}^2
\right),
\label{weyleuler} \ee
with $\beta=-(n_S+11 n_F+62n_V)/360=-N^2/4$.
The divergence of $\Gamma(2-(d+1)/2)$  in dimensional regularization 
in the limit $d+1\to 4$ corresponds to a $\log(1/\epsilon^2)$ divergence 
in the cutoff regularization we are using, with $\epsilon$ a length scale.
A comparison of the third term in eq. (\ref{effaction}) and the above 
expression 
reproduces the standard relation $G_5=\pi R^3/(2N^2)$ between
the bulk Newton's constant and the central charge of the dual CFT.
With this relation, the dimensionful UV momentum cutoff for $d=3$ can be expressed as
$\left( \epsilon_N R \right)^{-2}=2G_5/(R^3 G_4)=8\pi^2 m_{\rm Pl}^2/N^2$, with $m_{\rm Pl}^2=1/(8\pi G_4)$.
Now eq. (\ref{holentropydS}) for $d=3$ can be cast in the form 
\be
S_{\rm dS}
%=\frac{A_H}{4G_4}+\frac{R^3 S^2}{8G_5}\log(H\epsilon)
=\frac{A_H}{4G_4}+N^2 \log(H\epsilon_N)
=\frac{A_H}{4G_4}+N^2 \log \left( \frac{N}{\sqrt{8}\pi}\frac{H/R}{m_{\rm Pl}}
\right),
\label{holentropydSdual} \ee
where $H/R$ is the physical Hubble scale.
This expression is completely analogous to the black-hole result \cite{sen}, 
with the horizon size parameter measured in units 
of the UV cutoff. It is also in agreement with the calculation of the
logarithmic part of the holographic entanglement entropy in \cite{chm}. 

Our derivation of the dS entropy 
is consistent with the expectation that the entropy associated
with gravitational horizons can be understood as 
entanglement entropy if Newton’s constant
is induced by quantum fluctuations of matter fields \cite{jacobson}.
In the context of the AdS/CFT correspondence
the bulk degrees of freedom correspond to the matter fields of the 
dual theory. Within a construction that implements a UV cutoff, such as 
the RS model \cite{rs}, the Einstein action arises through the integration
of these bulk degrees of freedom. 
This is demonstrated explicitly by the expression (\ref{Geffrs}) for the effective 
Newton's constant. 
The leading contribution to the entropy has a universal
form that depends only on the horizon area, because the same degrees
of freedom contribute to the entropy and Newton's constant. Also, the
detailed nature of the 
UV cutoff does not affect the leading contribution. 
The particular features of the underlying theory, such as the number of 
degrees of freedom  
become apparent at the level of the subleading corrections to
the entropy. However, a level of universality still persists: 
the coefficient of the logarithmic correction is determined by the
central charge of the theory, and is independent of the regularization. 

It has been argued that the presence of a
UV cutoff is automatic even in theories in which gravity is not induced
\cite{dvali}. In particular, for a theory with a large number of 
degrees of freedom $N_{\rm dof}$, the UV momentum cutoff is expected to
be proportional to 
$m_{\rm Pl}/\sqrt{N_{\rm dof}}$. Our analysis reproduces this 
relation for a UV cutoff $1/(\epsilon_N R)$ and $N_{\rm dof}\sim N^2$. 
It seems likely then that the connection between entanglement and 
gravitational entropy is not limited to the case of induced gravity.

\section*{Acknowledgments}
I would like to thank C. Bachas, A. Belin, R. Brustein, G. Dvali, D. Giataganas, 
and E. Kiritsis for 
useful discussions. 
This research work was supported by the Hellenic Foundation for
Research and Innovation (H.F.R.I.) under the “First Call for H.F.R.I.
Research Projects to support Faculty members and Researchers and
the procurement of high-cost research equipment grant” (Project
Number: 824).

%\newpage

\end{document}